\documentclass[pra,aps,twocolumn,superscriptaddress,notitlepage]{revtex4-2}
\usepackage{amssymb,amsfonts,amsmath,amstext,graphicx,hyperref,lipsum,empheq}
\usepackage{float}
\usepackage[normalem]{ulem}
\hypersetup{
	colorlinks=true,
	linkcolor=blue,
	filecolor=magenta,      
	urlcolor=cyan,
}
\usepackage{amsthm}
\usepackage{outlines}
\usepackage{tikz}
\usetikzlibrary{shapes.geometric, arrows}

\tikzstyle{startstop} = [rectangle, rounded corners, minimum width=3cm, minimum height=1cm,text centered, draw=black, fill=red!30]
\tikzstyle{process} = [rectangle, minimum width=3cm, minimum height=1cm, text centered, draw=black, fill=blue!20]
\tikzstyle{decision} = [diamond, minimum width=3cm, minimum height=1cm, text centered, draw=black, fill=green!20]
\tikzstyle{arrow} = [thick,->,>=stealth]

\usepackage{outlines}
\usepackage{subfigure}
\usepackage{cleveref}
\usepackage{appendix}

\newcommand{\beq}[0]{\begin{equation}}
\newcommand{\eeq}[0]{\end{equation}}
\newcommand{\non}{\nonumber}
\def\be{\begin{equation}}
\def\ee{\end{equation}}
\def\bea{\begin{eqnarray}}
\def\eea{\end{eqnarray}}
\newcommand{\ba}{\begin{eqnarray}}
\newcommand{\ea}{\end{eqnarray}}
\usepackage{hyperref}

\def\BraVert{\egroup\,\mid\,\bgroup}

\definecolor{myblue}{rgb}{.8, .8, 1}

\begin{document}

\title{Generating discrete time crystals through optimal control}
\author{Mrutyunjaya Sahoo}%
\affiliation{Department of Physical Sciences, Indian Institute of Science Education and Research Berhampur, Berhampur 760010, India}
 \author{Rahul Ghosh }%
 \affiliation{Department of Physical Sciences, Indian Institute of Science Education and Research Berhampur, Berhampur 760010, India}
  \author{Bandita Das }%
 \affiliation{Department of Physical Sciences, Indian Institute of Science Education and Research Berhampur, Berhampur 760010, India}
  \author{Shishira Mahunta}%
 \affiliation{Department of Physical Sciences, Indian Institute of Science Education and Research Berhampur, Berhampur 760010, India}
 \author{Bodhaditya Santra}
 \affiliation{Department of Physics, Indian Institute of Technology Delhi, New Delhi, India}
\author{Victor Mukherjee}%
\affiliation{Department of Physical Sciences, Indian Institute of Science Education and Research Berhampur, Berhampur 760010, India}

\begin{abstract}
In this work we use optimal control to generate Discrete Time Crystals (DTC) in generic many-body quantum systems. We define appropriate cost functions, which, when optimized, result in the formation of DTCs. This hitherto unexplored method represents DTCs as an optimization problem, and allows us to find non-trivial realistic periodic control pulses and parameter regimes which result in spontaneous breaking of time-translational symmetry in quantum systems. We exemplify our approach using many-body quantum systems in the presence, as well as absence of dissipation. We also discuss possible experimental realization of the control protocol for generating DTCs.

\end{abstract}

\maketitle

{\it Introduction.--}Time crystals are a comparatively recently discovered non-equilibrium phase of matter, associated with spontaneous breaking of time translational symmetry \cite{sacha18time, else20discrete, sacha20time, zalatel23colloquium}. Discrete time crystals (DTCs) are formed in the presence of periodic modulation in time \cite{else16floquet}, while continuous time crystals are formed through the generation of limit cycles in the presence of dissipation \cite{iemini2018boundary, carollo24quantum}.  In the recent years, there has been an immense interest in finding ways to generate DTCs in various many-body quantum systems \cite{zalatel23colloquium}, and constructing the dynamical phase diagrams of many-body quantum systems under periodic modulation \cite{yao18discrete}. DTCs have been shown to exist in many-body localized systems \cite{yao18discrete, yao18time}, in the presence of Stark localization \cite{kshetrimayum20stark, liu23discrete} in clean quantum systems \cite{huang18clean, barlev24discrete} in the presence of long-range interactions \cite{russomanno17floquet, ho17critical, pizzi21higher}, in the presence of dissipation \cite{gong18dtc, lazarides20time, das24discrete}, and in non-Hermitian systems \cite{yousefjani25non}. DTCs have also been realized experimentally in ion trap \cite{zhang17bservation}, optical cavity \cite{kessler21observation, taheri22all} and nanoelectromechanical \cite{sarkar25observation} setups.  However, till now open questions remain regarding universal protocols  for generating DTCs in arbitrary many-body quantum systems. For example, analytically obtained control pulses which are suitable for generating DTCs are model dependent, and are usually restricted to simple forms, such as on-off \cite{gong18dtc, yao18discrete, else16floquet} or sinusoidal pulses \cite{jara24theory, Simon2024}. On the other hand, one can envisage practical scenarios, where varied constraints in experimental setups may preclude the implementation such simple modulation pulses.  Consequently, a universal protocol for generating DTCs in arbitrary many-body quantum systems is a crucial open question, which we address in this work. Notably, as we discus below,  this protocol may also assist in finding parameter regimes supporting DTCs in periodical modulated systems, which otherwise can be highly time-consuming in many-body quantum systems, specially in the presence of disorder \cite{yao18discrete}.

In this work we use optimal control of quantum systems, through Chopped RAndom Basis (CRAB) optimization protocol \cite{caneva11chopped, doria11optimal}, to generate DTCs in closed and open many-body quantum systems. We define appropriate cost functions, which when minimized through CRAB, result in spontaneous breaking of time-translational symmetry. This hitherto unexplored method presents us a universal way to generate DTCs in arbitrary many-body quantum systems, thereby allowing us to find non-trivial control pulses and parameter regimes for generating DTCs. Notably, optimal control of quantum systems have been highly successful for achieving various tasks in quantum condensed matter physics \cite{caneva09optimal, caneva11speeding}, and quantum technologies \cite{das23precision, ansel24optimal}, in both theoretical \cite{llyod14information, deffner17quantum}, and well as experimental settings \cite{borselli21two, vanfrank16optimal, muller22one}. However, to the best of our knowledge, the application of optimal control in the field of time crystals is still an unexplored topic. Here we bridge this gap, and show that optimal control can indeed be highly relevant for the development of DTCs as well. 
\label{secI}

{\it Methodology.--}DTCs are associated with long range orders both in space and time, and are characterized by an observable $\mathcal{O}(t)$ in a many-body quantum system, which when observed stroboscopically, varies periodically with a time period $ s  T$ ($s \in \mathbb{Z},~ s \geq 2$), in response to an external modulation of time period $T$. An equivalent description in the Fourier space of the observable $\mathcal{O}(t)$ corresponds to a single peak in the Fourier transformed amplitude at a frequency $\Omega_0 / s$ ($\Omega_0 = 2\pi /T$) \cite{yao18time}. Here we harness the above quantitative properties of DTCs to generate time translational symmetry breaking (TTSB) through optimal control. In particular, we consider a many-body system described by a Hamiltonian $H(\lambda_t)$, parametrized by a periodically modulated scalar parameter $\lambda_t = \lambda_{t + T}$ with a time period $T$. Following the CRAB scheme, $\lambda_t$ is expressed as a truncated Fourier series such that \cite{caneva11chopped}
\begin{equation}
     \lambda_t = A_0 + \frac{1}{2 N_c} \sum_{n=1}^{N_c}(A_n \cos(\nu_n t) +B_n \sin(\nu_n t)).
    \label{eq:Pulse}
 \end{equation}
Here $\nu_n = 2\pi n/T$, and the positive integer $N_c$ denotes the number of frequencies considered. We numerically optimize the Fourier coefficients $A_n$ and $B_n$ so as to minimize an appropriately chosen cost function $\mathcal{F}(\{ A_n, B_n \})$, subject to certain physically relevant constraints, in order to generate a DTC. Without loss of generality, here we choose $s = 2$.

In order to perform the above mentioned optimization, we start with an initial guess pulse $\lambda_t(\{A^{in}_n, B^{in}_n\})$. We then apply the CRAB optimization scheme, subject to the constraint $|A_n|, |B_n| \leq \chi,~\forall~n$, where the positive constant $\chi$ denotes a bound on $A_n,~B_n$. We expect the control protocol will iteratively generate an optimal pulse $\lambda_t(\{A^{opt}_n, B^{opt}_n\})$, which results in a DTC, if the model and the parameter values allow for the existence of the same. \\

 We apply the above protocol in many-body models both in the presence as well as absence of dissipation, to show that optimal control may generate DTCs in varied many-body quantum systems. We note that an equilibrium phase corresponds to minimum of the free energy, and is inherently related to equilibrium phase transitions driven by thermal fluctuations. In contrast, here  minimization of the relevant cost function can be expected to give rise to DTC phase, for suitable values of the parameters describing the system.  
 The overall method is depicted as follows: {\it define the appropriate cost function $\to$ start with an initial guess pulse $\{A_{n}, B_n \} = \{A_{n}^{in}, B_n^{in} \}$ $\to$ find the coefficients $\{A_{n}, B_n \}$ so as to minimize the cost function $\to$ the final optimal pulse generates a DTC}. We now exemplify the above protocol using open and closed many-body quantum systems.

{\it DTC in the modulated open Dicke model.--} We consider a modulated open Dicke model comprising  $N > 1$ spins $1/2$'s collectively coupled to a photonic mode, described by the following Hamiltonian 
\begin{align}
\hat{H}_{OD}(\lambda(t)) &= \omega a^\dagger a + \omega_0 J_z + \frac{2\lambda_t}{\sqrt{N}}(a^\dagger  +a) J_x. \label{eqDicke}
\end{align}
Here  $\omega  = \omega_T(1-\epsilon)$ and $\omega_0 = \omega_T(1+\epsilon)$,  where $\epsilon$ denotes the  detuning between the frequencies $\omega$ and $\omega_0$, $\omega_T > 0$, $ J_\mu \equiv \frac{1}{2} \sum_i^N \hat{\sigma}_i^\mu  $ denotes the angular momentum operator acting on the spins along $\mu = x, y, z$ direction, $a$ ($a^{\dagger}$) denotes the photonic annihilation (creation) operator,  and $\lambda_t$ is the periodically modulated coupling strength between the spins and the photonic mode. We consider a leaky cavity, such that the Lindblad master equation describes the dynamics of the whole system:
\begin{equation}
\frac{d \hat{\rho}_t}{dt} =  -i [\hat{{H}}_{OD}, \hat{\rho}_t] + \kappa D[\hat{a}] \hat{\rho}_t.
\end{equation}
Here $\kappa > 0$ denotes the rate of dissipation of the photons into the vacuum, and $D[\hat{a}]\hat{\rho} \equiv \hat{a} \hat{\rho} \hat{a}^{\dagger} - \frac{1}{2}\left(\hat{a}^{\dagger} \hat{a} \hat{\rho} + \hat{\rho} \hat{a}^{\dagger} \hat{a}   \right)$. In the thermodynamic limit $N \to \infty$ the system goes through a  $\mathbb{Z}_2$  symmetry-breaking phase transition at $\lambda = \lambda_c = \sqrt{\frac{\omega_0}{\omega} \left( \omega^2 + \frac{\kappa^2}{4} \right)}$. One can use the mean-field approximation to evaluate the dynamics of the setup in the thermodynamic limit $N \to \infty$ \cite{gong18dtc}.

We get two symmetry-broken steady states for $\lambda > \lambda_c$, given by:
\ba
(j_x^\pm, j_y^\pm, j_z^\pm) &= \frac{1}{2}(\pm \sqrt{1 - \mu^2}, 0, -\mu) \non\\
(x_\pm, p_\pm) &= \mp \left[\frac{\sqrt{2\omega(1 - \mu^2)}}{\omega^2 + \kappa^2/4}\right](\lambda, \kappa/2)
\label{eqsteadystates}
\ea
whereas there is a unique steady state for $\lambda < \lambda_c$: $(j_x,j_y,x,p) =(0,0,0,0) , j_z = \frac{1}{2}$ with $\mu = \frac{\lambda_c^2}{\lambda^2}$ \cite{gong18dtc}. 

In Ref. \cite{gong18dtc}, the authors used Markovian dynamics to show that starting from one of the symmetry-broken steady states, one can periodically modulate the parameter $\lambda$ in presence of a detuning $\epsilon$, in order to generate a DTC for weak dissipation; the same was obtained for intermediate values of dissipation in the presence of non-Markovian dynamics in Ref. \cite{das24discrete}. In the above works, the authors used a $\lambda_t$ of the form 

 \begin{equation}
     \lambda_{t+T} = \lambda_t = 
     \begin{cases} 
     \lambda_0 & \text{for } 0 \leq t < \frac{T}{2}, \\
     0 & \text{for } \frac{T}{2} \leq t < T.
     \end{cases}
 \end{equation} \\
to generate the DTCs.  However, one can envisage an experimental setup, in which, owing to different constraints, such kind of "on-off" pulses are impossible to implement. As such, the presence of a protocol which can lead us to more generic DTC-generating pulses, subject to experimental constraints, can be of immense importance. As we discuss below, the CRAB protocol described here can inherently incorporate these experimental constraints, and yet, generate pulses which can lead to the formation of DTCs.

\begin{figure}[h] 
     \centering 
    \includegraphics[width=0.4\textwidth, height = 0.55\textwidth]{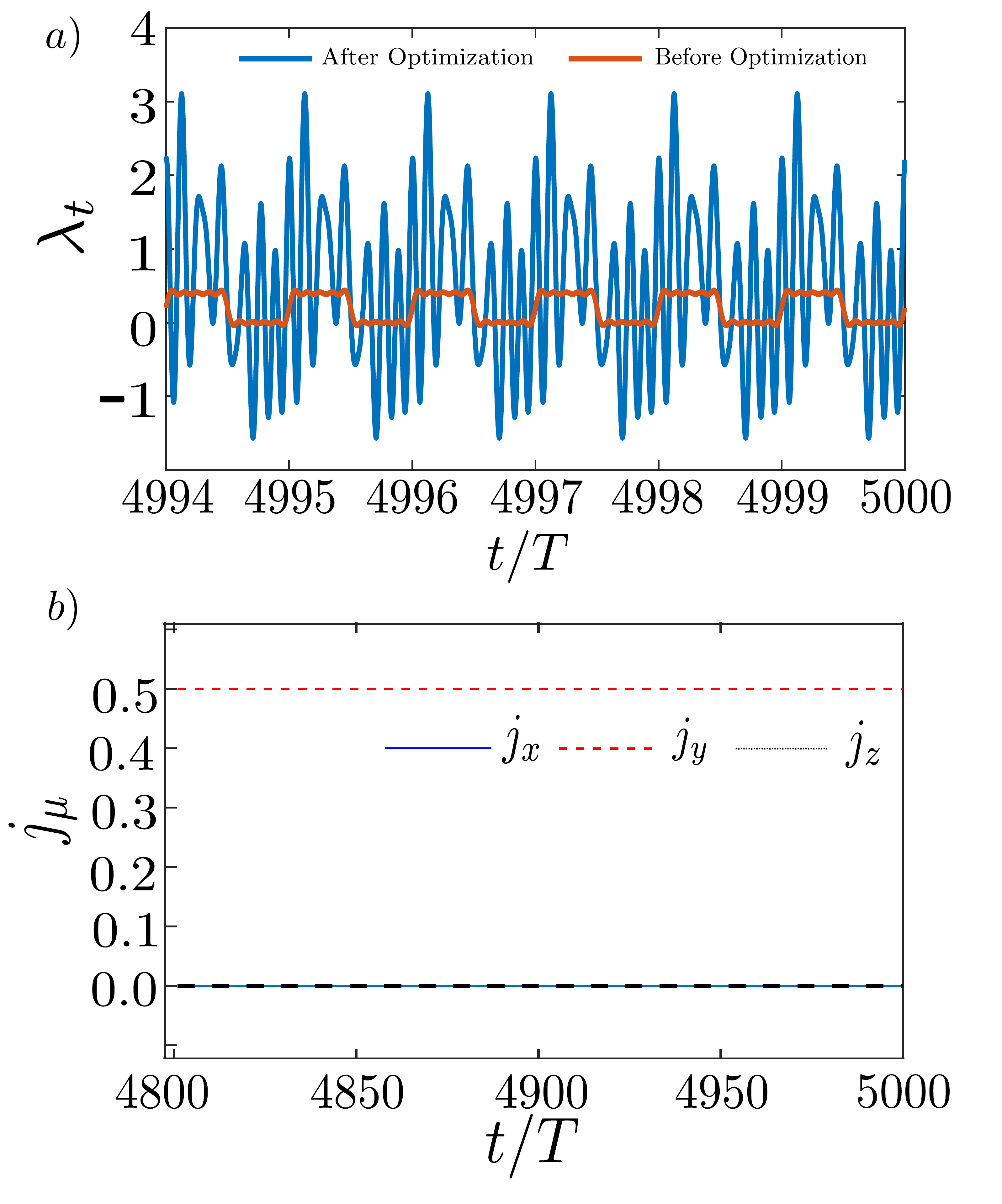} 
    \caption{Plot (a) shows the initial guess pulse and the final optimised pulse, respectively, for the open Dicke model. Plot (b) shows the stroboscopic dynamics for the initial guess pulse. Here $\epsilon = 0.05$, $\chi=10$, $N_c=10$.}
    \label{fig:Dickemodelbeforeandafteropt} 
 \end{figure}

\begin{figure}[h] 
     \centering 
    \includegraphics[width=0.53\textwidth,  height = 0.75\textwidth]{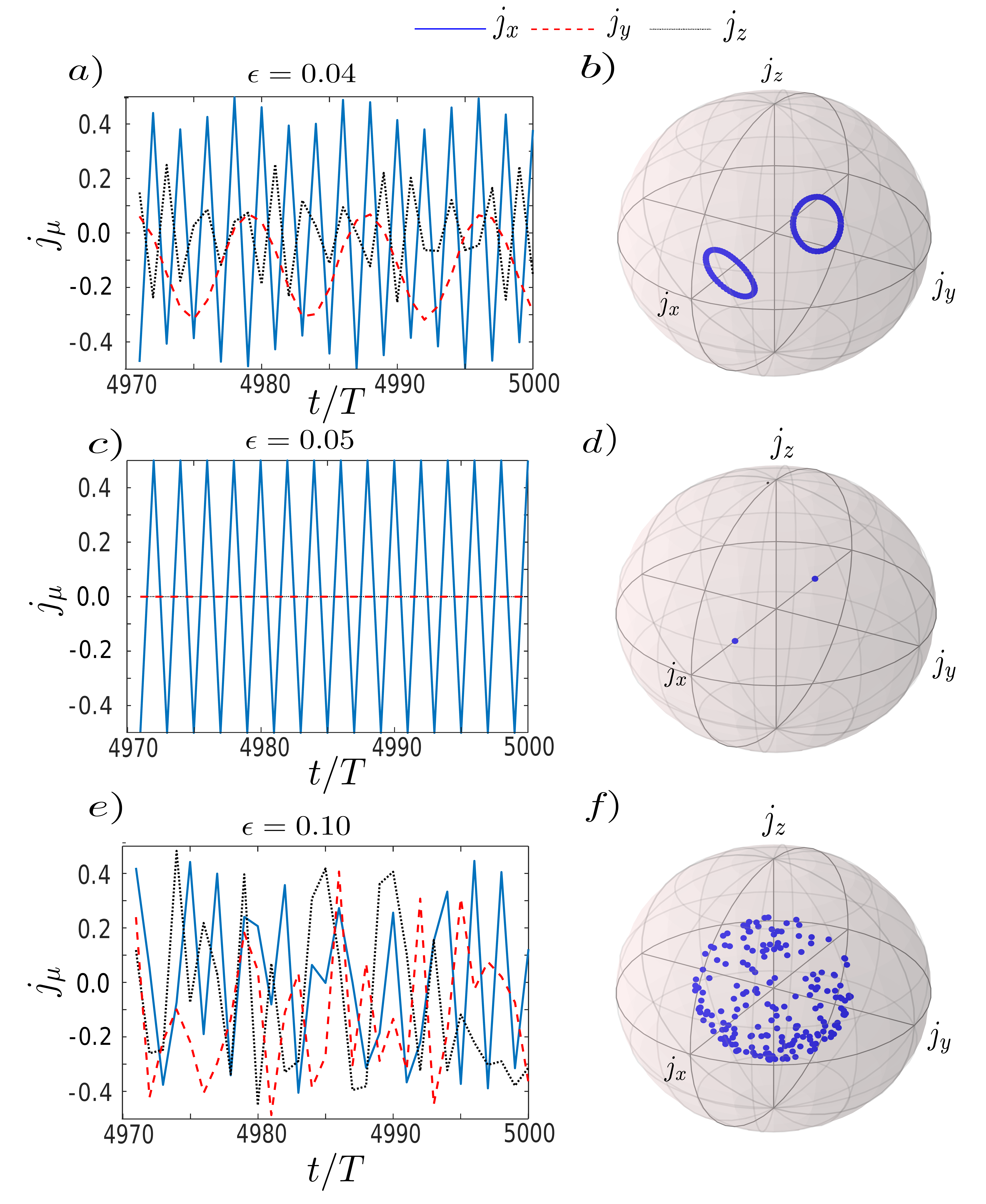} 
    \caption{We show the behaviour of $j_{\alpha}(t)$ ($\alpha = x, y, z$) for different values of the detuning $\epsilon$ in  (a), (c), (e) and the corresponding Bloch sphere pictures in (b), (d) and (f) for the optimized pulse shown in Fig. (\ref{fig:Dickemodelbeforeandafteropt} a (in blue)), for the open Dicke model. As we change $\epsilon$, we get different dynamical phases, viz. limit cycle for $\epsilon = 0.04$ in (a) and (b); DTC for $\epsilon  = 0.05$ in (c) and (d); and a thermal phase for $\epsilon = 0.1$ in (e) and (f). Here $\kappa = 0.05$.}
    \label{fig:DynamicDickeModel} 
 \end{figure}

In order to generate a DTC in the open Dicke model, we consider a periodic modulation of the form Eq. \eqref{eq:Pulse}. We carry out the optimization such that the cost function
\ba
\mathcal{F}_1(\{A_n,B_n\}) &=& \left| j_x(sT) - j_x(sT+2T) \right| \non\\ &+&  \frac{1}{\left| j_x(sT) - j_x(sT+T) \right|}
\label{eq:costfunction1}
\ea
is minimized, for some integer $s \gg 1$, and some choice of the detuning parameter $\epsilon$ (see Eq. \eqref{eqDicke}). 
In the above equation \eqref{eq:costfunction1}, the first term on the r.h.s. ensures that as a result of the optimization, $j_x(t)$ becomes $2T$ periodic, whereas the second term on the r.h.s. denotes a penalty for pulses which result in  $T$ periodic $j_x$'s.  We start with a guess pulse, characterized by an initial choice of the Fourier coefficients $A_n^{in}$ and $B_n^{in}$, which in general will not result in a DTC (see Fig. \ref{fig:Dickemodelbeforeandafteropt}).
However, as we show in Figs. \ref{fig:Dickemodelbeforeandafteropt} and \ref{fig:DynamicDickeModel}, the above described optimization protocol finally results in a DTC generating pulse. We further check the robustness of the DTC resulting from the optimal pulse, by changing the strength of the detuning parameter $\epsilon$.  As shown in Fig. \ref{fig:DynamicDickeModel}, the DTC phase persists for a finite range of $\epsilon$, in addition to the existence of other dynamical regimes and thermal phases.

\begin{figure}[h] 
    \centering
    \includegraphics[width=0.5\textwidth, height = 0.32\textwidth]{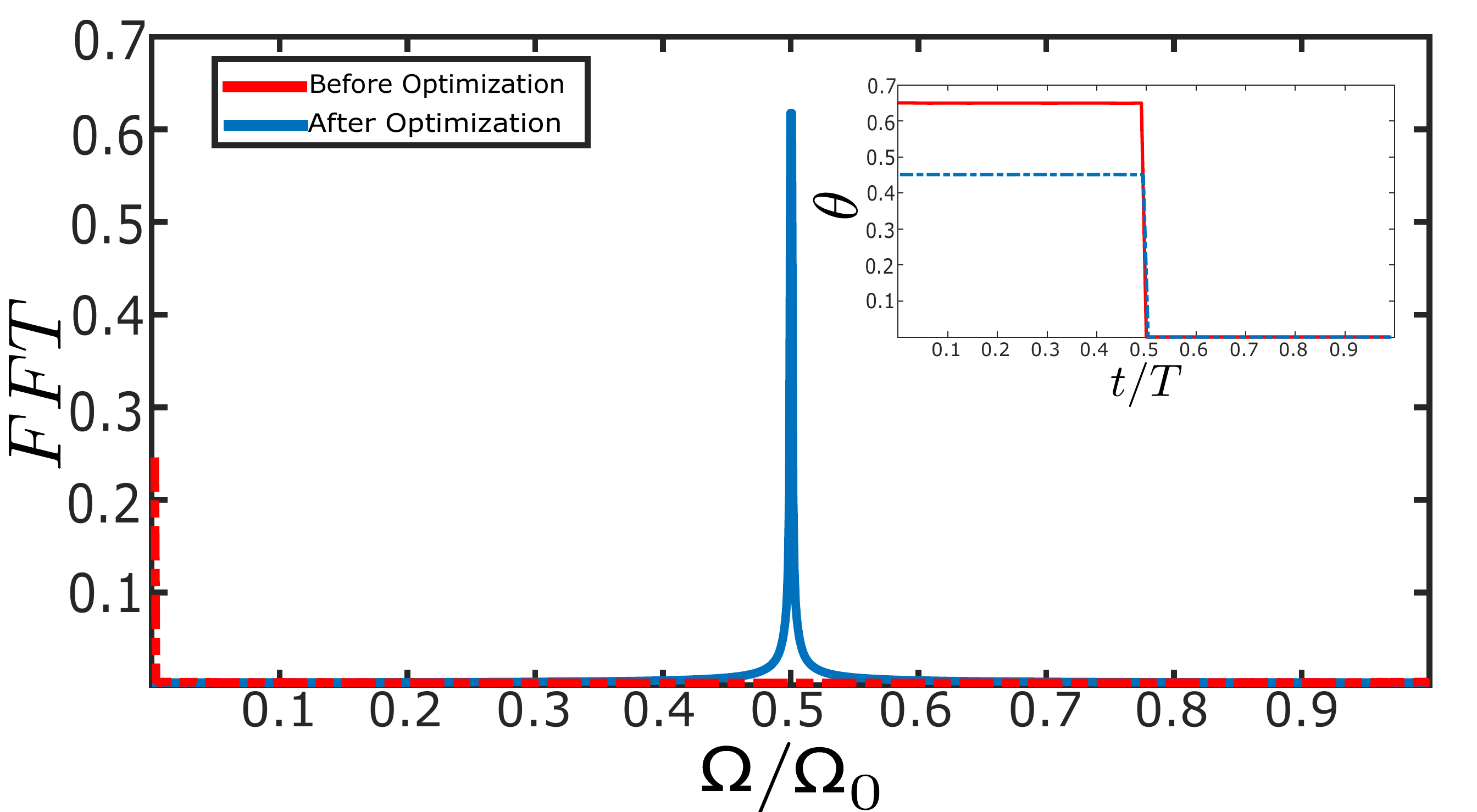} 
    \caption{We show the $\theta$ before and after optimisation for the cost function (see Eq. \ref{cfgenfourier}), along with the corresponding resultant FFT for the spin chain model \eqref{eq:MBL}.
        The spin chain length is taken to be $L=8$. 
        To perform the optimization, the cost function (\ref{cfgenfourier}) is evaluated by averaging over the lattice sites $i=4, 5$ and all possible initial product states. We impose the constraints $|A_0| < 1$ and $|A_n|, |B_n| < 0.0005$ for $n = 1,2, 3, \dots, N_c = 6$.}
    \label{fig:MBL_Spin_Chain}
\end{figure}

{\it DTC in a many-body-localized spin chain system.--}
In order to verify the applicability the above introduced optimization scheme in closed quantum systems and in the presence of disorder, we next consider a spin chain described by a periodically modulated Hamiltonian $H(t) = H(t + T)$ of the form \cite{yao18discrete}
\begin{equation}
        H(t) =
        \begin{cases} 
        H_1 \equiv (g - \theta) \sum_i \sigma_x^i,  &\text{for} ~0 < t \leq T_1 \\
        H_2 \equiv \sum_i J_z^i \sigma_z^i \sigma_z^{i+1} + \sum_i B_z^i \sigma_z^i,  &\text{for} ~T_1 \leq t < T.
        \end{cases}
        \label{eq:MBL}
\end{equation}
In the above equation \eqref{eq:MBL}, $J_z^i$  denotes the interaction between the spins at sites $i$ and $i+1$ for the time interval 
$T_1 \leq t < T$, $B_z^i$ denotes a longitudinal field acting on the site $i$ for the same time interval, while $H_1$ describes a transverse field that acts for the time interval $0 < t \leq T_1$. For $\theta = 0$ and $gT_1 = \pi/2$, $H_1$ describes a perfect spin flipping field. On the other hand, in general in the presence of a non-zero $\theta$, the spins can be expected to go out of phase, thereby resulting in a thermal state after a few modulation periods $T$. However, as shown  in \cite{yao18discrete}, the disorder present in $H_2$ can lead to many-body localization. This in turn can result in a DTC phase in the presence of the periodic modulation \eqref{eq:MBL}, characterized by a periodically varying $\langle \sigma_z(t) \rangle$ with a time period $2T$, which is robust to small $\theta$ for suitable values of the Hamiltonian parameters. Alternatively, the DTC phase can also be characterized by the fourier transform (FFT) of the auto-correlation function 
$R_i(t) = \frac{1}{2^L} \sum_{\{z\}} \langle z|  \sigma^i_z(t) \sigma^i_z(0) | z\rangle$, where the sum is over all possible $2^L$ product states $|z\rangle$ for a spin-chain of length $L$;
$R_i(t)$   shows a delta function peak at a frequency $\Omega = \Omega_0/2 = \pi/T$ in the DTC phase.

In general, finding a characteristic value of $\theta$ which can result in a DTC phase for a given set of parameter values $\{J_z^i, B_z^i\}$ may require multiple trial and errors, which can be a highly non-trivial and time-intensive task, specially for large system sizes in the presence of disorder. In contrast,  here we show that the protocol developed in this work can be immensely helpful for finding a suitable regime of $\theta$ which can result in a DTC phase, thereby simplifying reproduction of the dynamical phase diagram considerably. 

To this end, we consider the modulation parameter $\theta$ to be time-dependent, which needs to be optimized. The optimal $\theta = \theta_{opt}$ hence obtained will give us an order of magnitude of $\theta$ such that the DTC phase is preserved for $|\theta| \lesssim |\theta_{opt}|$.  In particular, we choose
\begin{equation}
    \small
    \theta_{t} = \theta_{t + T} =
    \begin{cases} f(t)
     &  0 \leq t < \frac{T}{2} \\
    0 &  \frac{T}{2} \leq t < T.
    \end{cases}
    \label{detuningpulseCRABMBL}
\end{equation}
where as before, $f(t)=A_0 + \frac{1}{2N_c}\sum_{n=1}^{N_c}(A_n \cos(\nu_n t) +B_n \sin(\nu_n t))$.
For simplicity we have taken $T_1 = T/2$. In general the above choice of modulation results in a $\theta_{opt}$  that is time dependent for $0 < t \leq T/2$, while imposing a constraint $0 < \{|A_n|, |B_n|\}_{\rm max} \ll |A_0|$ will result in the $\theta_{opt}$ to assume an almost constant but non-zero value for that time-interval.
In order to ensure that the optimization scheme results in a FFT which peaks at $\Omega = \Omega_0/2$, we consider the following cost function to be minimized:
\ba 
    \small
    \mathcal{F}_2(\{A_n, B_n\}) &=& \mathcal{F}_2^{(1)} (\{A_n, B_n\}) + \mathcal{F}_2^{(2)} (\{A_n, B_n\}) \non \\
    &+& \mathcal{F}_2^{(3)} (\{A_n, B_n\}), \non\\
    \mathcal{F}_2^{(1)} (\{A_n, B_n\}) &=& |FCMA-0.5\Omega_0|,\non\\
    \mathcal{F}_2^{(2)} (\{A_n, B_n\}) &=& \sum_{\omega \neq 0.5 \Omega_0\ }|FFT(\omega)|,\non\\
    \mathcal{F}_2^{(3)} (\{A_n, B_n\}) &=& { \frac{\Theta(x)}{\left|x\right|}};~~x = FFT(\Omega_0/2)-0.05.
     \label{cfgenfourier}   
\ea
Here FCMA denotes Frequency Corresponding to Maximum Amplitude of FFT, while
\ba
     \Theta(x) &=& 
    \begin{cases} 
    10^{4} 
     & \text{for }  x  \leq 0, \\
    1 & \text{for }  x> 0.\non\\
    \end{cases}
\ea
The first ($\mathcal{F}_2^{(1)}$)  and second  ($\mathcal{F}_2^{(2)}$) parts in the r.h.s of Eq. ~\eqref{cfgenfourier} ensure that the peak occurs at half of the driving frequency and no other peaks exist at any other frequencies, respectively, whereas, the third part ($\mathcal{F}_2^{(3)}$) defines a lower threshold value ($= 0.05$) for the FFT peak at $\Omega = \Omega_0/2$, for the phase to be considered as a DTC \cite{yao18discrete}. 
Here $gT =\pi$, the amplitudes $J_z^i$ are chosen from a uniform sampling of $J_z^i \in [J_z-0.2 J_z,J_z+0.2 J_z]$, where $J_z = 0.2\pi$, and the on-site potentials are chosen from a uniform sampling of $B^i_z \in [0,2\pi]$.

As shown in Fig. \ref{fig:MBL_Spin_Chain}, we start from an initial guess pulse with a large $\theta=0.65$ , which does not correspond to a DTC phase. In systems involving a large number of lattice sites, in most practical scenarios, it might be feasible to evaluate the autocorrelation function for a small subsystem only. Consequently, here we consider FFT averaged over the spin sites $i = 4$ and $i = 5$ only for implementing the optimization protocol. As expected, the optimization scheme iteratively takes us to a final smaller $\theta = \theta_{opt} = 0.45$, which generates a DTC, as shown by a delta function form of the FFT. We emphasize that even though here the optimization is carried out by considering a small central subsystem ($i = 4, 5$), the resultant optimal pulse generates a global DTC phase, as verified by considering the FFT of the autocorrelation function averaged over all the lattice sites $\bar{R}(t) = \frac{1}{L} \sum_{i=1}^L R_i(t)$ (see Fig. \ref{fig:MBL_Spin_Chain}).

{\it Possible experimental realization.--}
Here we describe a way to experimentally realize  the symmetry-broken steady states for $\lambda > \lambda_c$, given by Eq. \ref{eqsteadystates}, which supports the observation of DTC through optimal control. The paradigmatic scenario of highly directional optical emission from incoherently excited two level atoms were discussed by Dicke. The phenomena was observed in the form of Rayleigh scattering off a Bose-Einstein condensate excited by a single off resonant laser beam \cite{Inouye1999}. In this experiment, the interference between recoiling atoms and the condensate formed a matter-wave grating, which enhanced subsequent light scattering in specific directions, demonstrating a positive feedback mechanism characteristic of superradiance. The process resulted in a burst-like emission of photons and atoms, showcasing coherent collective dynamics in a dilute gas system.

To realize the steady states described by equation (\ref{eqsteadystates}) and enable the observation of discrete time crystal dynamics through optimal control, we propose an experimental scheme involving a Bose-Einstein condensate of ultracold $^{87}$Rb atoms coupled to an optical cavity. The atoms can be prepared in a well-defined hyperfine ground state, such as 5$^2$S$_{1/2}$ $|F=2,m_F=2⟩$, and confined in a magnetic or hybrid trap that produces a cigar-shaped condensate elongated along the cavity axis.

The BEC is aligned with the axis of the high-finesse optical cavity, while a far-detuned pump beam is applied perpendicular to the cavity axis, forming a transverse optical standing wave. This configuration facilitates dispersive atom-light coupling, with negligible spontaneous emission, and allows for selective momentum transfer between atomic states via photon scattering. The cavity linewidth is assumed to be narrow enough to enable momentum-resolved processes, similar to those demonstrated in previous superradiant and collective scattering experiments \cite{Baumann2010, Kessler2021, Kindler2015}.

Atomic populations in different momentum states can be monitored through time-of-flight imaging or cavity transmission, enabling the detection of subharmonic responses characteristic of DTC behavior. The parameters of the pump field and the trap can be tuned to match the desired driving protocols, and optimal control methods may be applied to modulate the temporal structure of the drive, enhancing stability and coherence of the DTC phase. This setup combines well-established elements - cavity QED with BECs, dispersive light-matter interaction, and momentum-resolved atom detection — making it a viable platform to realize and probe discrete time crystalline behavior using optimized pump field in state-of-the-art experimental setups. 

Another platform for realizing DTC with optimized pulses is spin chains composed of trapped atomic ions \cite{yao18discrete} In such systems time translation symmetry can be spontaneously broken when the system is in a non-equilibrium state such as many body localization. The spins can be encoded in the 6$^2$S$_{1/2}$ $|F=0,m_F=0⟩$ and $|F=1,m_F=0⟩$ hyperfine clock states of an electrostatically trapped and laser cooled $^{171}$Yb$^+$ ions \cite{zhang17bservation}. The many body localization hamiltonian can be realized using long-range Ising interactions and disordered local effective fields engineered by large AC Stark shifts \cite{Lee2016, Smith2016}.

{\it Conclusion.--}
In this work we have presented an optimization protocol which can generate DTCs in generic many-body quantum systems. We have exemplified our approach by applying the CRAB optimization scheme in the open Dicke model in the presence of dissipation, as well as in a disordered spin chain in the absence of dissipation. In either case, we show that optimal control generates robust DTC phases. Notably, in case of spin chain, optimization over a small central subsystem (lattice sites $i = 4, 5$) suffices to result in a pulse which generates a DTC, even when the complete spin chain is considered. 

We have then discussed possible experimental realization of DTC in an open Dicke model using CRAB optimization scheme, in a BEC setup, and in spin chains, using trapped ions. In addition, we note that
Dicke model has already been realized experimentally in optical cavity systems \cite{Baumann2010}; in the last few years several groups have experimentally shown the formation of DTCs using optical cavity setups \cite{Kessler2021, taheri22all}, ion traps \cite{zhang17bservation} and superconducting qubits \cite{frey22realization}. Similarly, CRAB optimization has been used in experimental works involving cold atom setups for varied tasks, including for controlling the dynamics of atoms \cite{rosi13fast}, for realizing time-reversal procedures \cite{mastroserio22experimental}, and for optimal preparation of quantum states \cite{lovecchio16optimal, borselli21two}. Consequently, 
we expect the control protocol presented here can be experimentally realized in several already existing platforms, and be relevant for near-term quantum technologies. 

\acknowledgements
B.D. acknowledges support from Prime Minister Research
Fellowship (PMRF). V.M. acknowledges support from Science and Engineering Research Board (SERB) through MATRICS (Project No.
MTR/2021/000055) and a Seed Grant from IISER Berhampur. \\

\appendix

\section{Robustness of DTC phase in the open Dicke model}

\begin{figure}[h] 
     \centering 
    \includegraphics[width=0.5\textwidth]{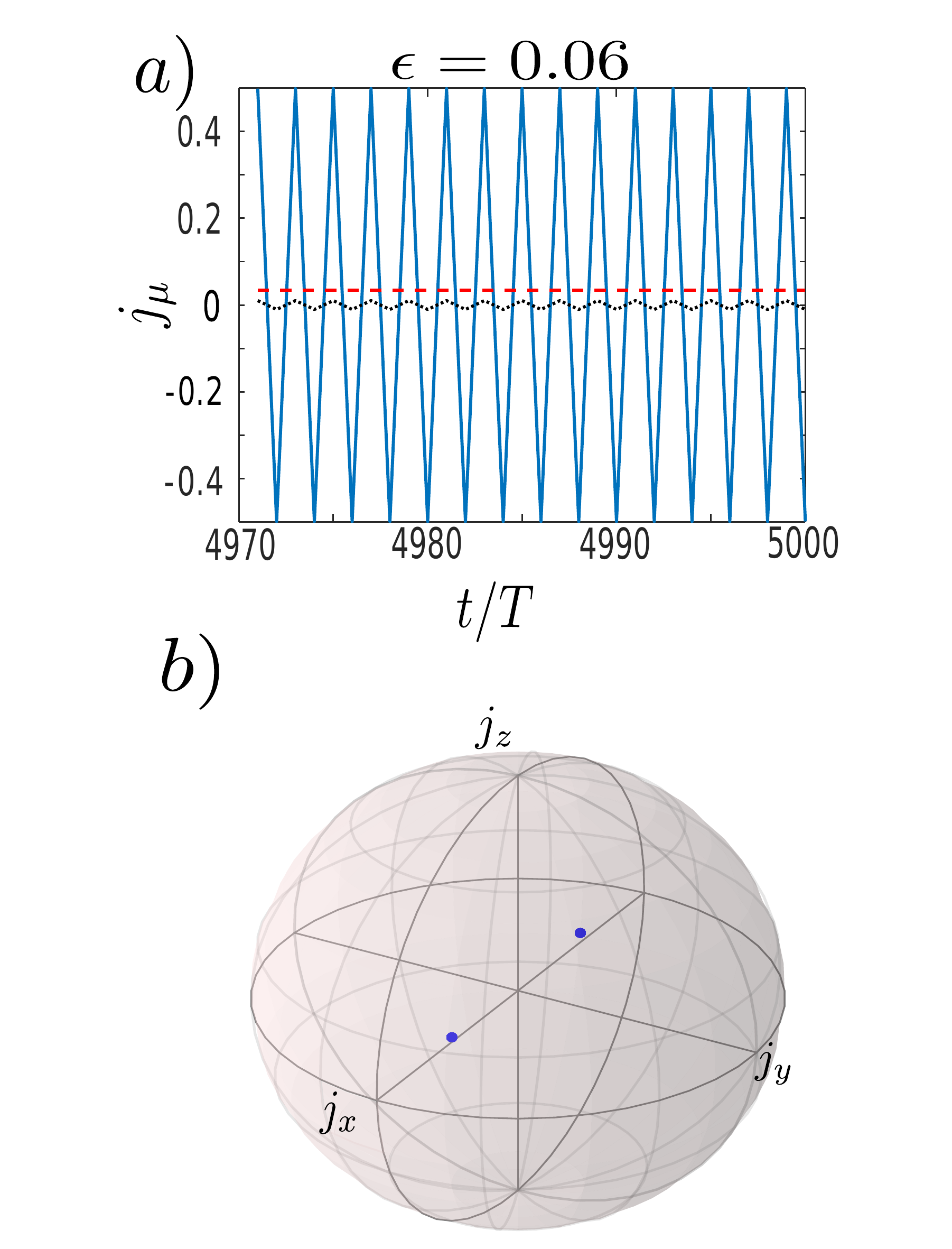} 
    \caption{Plot (a) shows the DTC signature for $\epsilon=0.06$ for the open Dicke model, and the corresponding Bloch sphere presentation is shown in (b).}
    \label{fig:Dickemodelbeforeandafteropteps006} 
 \end{figure}
In the open Dicke model, the optimization is carried out such that the optimal pulse generates a DTC for $\epsilon = 0.05$. However, as shown in Fig. \ref{fig:Dickemodelbeforeandafteropteps006}, the DTC phase persists even for $\epsilon = 0.06$, thus showing the robustness of the DTC phase over an extended regime of $\epsilon$.

\end{document}